\documentstyle[12pt]{article}

\def\gta{\mathrel{\mathpalette\fun >}}

\def\fun#1#2{\lower3.6pt\vbox{\baselineskip0pt\lineskip.9pt
  \ialign{$\mathsurround=0pt#1\hfil##\hfil$\crcr#2\crcr\sim\crcr}}}

\begin{document}

\title{Photon Statistics Limits for Earth-Based Parallax Measurements
of MACHO Events}
\author{Daniel E. Holz and Robert M. Wald \\
Enrico Fermi Institute and Department of Physics \\
University of Chicago \\
5640 S. Ellis Ave., Chicago, IL 60637-1433}
\date{}

\maketitle

\begin{abstract}
We analyze the limitations imposed by photon counting statistics on
extracting useful information about MACHOs from Earth-based
parallax observations of microlensing events. We find that if one or more
large (say $2.5\,\rm m$)
telescopes are dedicated to observing a MACHO event for several
nights near maximum amplification, then it is possible, in
principle, to measure the velocity of the MACHO well enough to distinguish
between disk and halo populations for events with
$\omega {A_m}\!^2 \gta 1\,\rm day^{-1}$, where $\omega^{-1}$
denotes the timescale
of the event and $A_m$ denotes its
maximum amplification. Thus, if it turns out to be possible to reduce
all other sources of error to below that of photon statistics, it may be
feasible to do useful Earth-based parallax observations for high
amplification events.
\end{abstract}

There are now a number of experimental efforts underway to detect
microlensing by massive compact halo objects (MACHOs).~\cite{MACHO,EROS,OGLE}
To date
there have been over 50 events reported. Events are now being
discovered in realtime,~\cite{EWS2,macho1} and telescopes can be
trained on events in the process of enfolding.

It is of considerable interest to determine the masses and velocities
of the MACHOs, since from the masses one could determine their
contribution to the dark matter in our galaxy, and from their velocities
one could determine whether they belong to a disk or halo population.
Unfortunately these parameters are not, in principle, determinable by
observations from a single telescope undergoing inertial motion. This
can be seen as follows:
the amplification of a lensed star
(which is all that can be measured) is given by
\begin{equation}
A(u)={u^2+2\over u(u^2+4)^{1/2}},
\label{Acurve}
\end{equation}
with $u=d/r_e$, where $d$ is the perpendicular distance
of the MACHO from the line joining the observer to the
star. Here $r_e$ is the Einstein radius, defined by
\begin{equation}
r_e=\sqrt{4 G m x (L-x)\over c^2 L},
\label{eradius}
\end{equation}
where $m$ is the mass of the lensing object, $L$ the distance
from the observer to the star, and $x$ the distance from the
observer to the lens. For inertial motions of the star, MACHO, and
observer, the time variation of $u$ is given by
\begin{equation}
u(t)=
\sqrt{\beta^2+\omega^2(t-t_0)^2},
\label{ueq}
\end{equation}
with $t_0$ the central time,
${\mbox{\boldmath $\omega$}}={\bf v}/r_e$, and
${\mbox{\boldmath $\beta$}}={\bf b}/r_e$,
where ${\bf v}$ is the velocity of the MACHO with
respect to the line between the observer and the star,
and $\bf b$ is its impact parameter
(so that ${\mbox{\boldmath $\omega$}}\perp
{\mbox{\boldmath $\beta$}}$).
Consequently, the light curve of a MACHO event is completely
characterized by the three parameters ($A_m$, $\omega$, $t_0$),
where $A_m$ is the maximum amplification, given in terms of $\beta$
by,
\begin{equation}
A_m={\beta^2+2\over\beta(\beta^2+4)^{1/2}}.
\end{equation}
It follows that the only information about $\bf v$ and $m$
that can be extracted directly from the light curve is the
value of the single parameter $\omega=v/r_e$.

The situation improves considerably, however, if an additional
telescope (widely separated from the first) also observes the
event.~\cite{Refsdal,Gould0,Gould1}
(Alternatively, similar improvement can occur if a single telescope
undergoes significant deviations from straight line motion over the
timescale of the event.~\cite{Gould0,Gould2})
As a result of parallax effects, the second
telescope will see a similar light curve, with the same value of
$\omega$, but with slightly different values of $A_m$ and $t_0$.  As
we shall see explicitly below, from this information one can, in
principle,~\footnote{
As analyzed in~\cite{Gould1}, there is a 2-fold
degeneracy in the determination of $\tilde r_e$, as it is possible for
the trajectory of the line between the lensed star and MACHO to pass
between the telescopes. However, the probability of this happening for
the short baseline observations of interest here is completely negligible,
and thus will be ignored in our analysis. (In any case, the degeneracy can
be resolved by using three telescopes or two telescopes
undergoing non-inertial motion.~\cite{Gould3})}
determine the ``reduced Einstein radius''
\begin{equation}
\tilde{r_e}={L\over(L-x)}r_e.
\end{equation}
The magnitude of the ``reduced transverse velocity''
($\tilde{\bf v}\equiv{L\over(L-x)}{\bf v}$)
is then given by
\begin{equation}
\tilde{v}={L\over(L-x)}r_e{\omega}
=\tilde r_e{\omega}.
\end{equation}
The parameters $\tilde r_e$ and $\tilde v$ yield considerable
further information about $m$ and $v$, although one must still
know or estimate the value of $x/L$ from other arguments to
determine $m$ and $v$ exactly.

For separations as large as $1\,\rm Au$ there should be no difficulty in
measuring the above parallax effect. Indeed, parallax effects due to
the Earth's orbit about the Sun have recently been
observed in an event lasting over two hundred days (the longest
event yet detected).~\cite{macho1} However, in order to make use
of the Earth's orbital motion, the lensing event
must span a fair portion of the earth's orbit.
Most events observed thus far have considerably shorter
timescales ($<100$ days), and therefore do not lend
themselves to this type of measurement.
With an additional telescope placed in solar orbit, however,
it should
be possible to measure parallax
effects for short timescale
events without difficulty, as has been
proposed by Gould.~\cite{Gould1,Gould3}

Unless one undertakes the major project of putting a telescope into
orbit, the effective baseline for a parallax observation of a short
timescale event (achieved either via an additional telescope or via the
rotation of the earth) is limited to about an earth radius. It has
generally been assumed that the very small changes to the light curve
associated with such a short baseline would be insufficient to allow
any useful parallax information to be extracted. However, the only truly
fundamental limitation on the ability to determine the parameters of
a light curve arises from ``photon statistics", i.e., the random
$\sqrt{N}$ fluctuations in the number of photons reaching a
telescope. In this paper, we investigate the limitations imposed by
photon statistics on parallax observations using
Earth-based telescopes.  Our conclusion is that although these
limitations are formidable, it is possible in principle to extract
useful information for
high amplification events.  Specifically, with one or more large (say
$2.5$ m) telescopes observing for several full nights,
it should be possible to determine $\tilde v$ well enough to
distinguish between disk and halo populations for events with
$\omega {A_m}\!^2$ greater than about $1\,\rm day^{-1}$. Thus, if all other
sources of error can be adequately controlled, it may be possible to
obtain some new, valuable information about MACHOs using Earth-based
astrometry.

The limitations imposed by photon statistics can be estimated by
analyzing the following two idealized problems for a single inertial
telescope which makes perfect measurements of the photon flux reaching
it: (i) Suppose the correct values of the parameters $\omega$ and
$A_m$ of the light curve for a MACHO event are given. What error would
photon statistics be expected to produce in the determination of $t_0$
from observations of the light curve? (ii) Suppose, similarly, that
$\omega$ and $t_0$ are given. What error would photon statistics be
expected to produce in the determination of $A_m$?

To answer the first question, we imagine ``binning" the photons which
arrive at the telescope into intervals of time $T$. The expected
photon flux in the bin at time $t$ is then $N=f_0A(t)T$, where $f_0$
denotes the average photon flux from the unamplified star and $A(t)$
denotes the amplification produced by the MACHO at time $t$. In
principle the central time, $t_0$, could be obtained from a
measurement of $N$ in a single bin (say the $i$th bin)
by inverting this equation, using
eqs.~\ref{Acurve} and~\ref{ueq} together with the given values of
$\omega$ and $A_m$. However, photon statistics produces an error
$(\Delta N)_i=\sqrt N_i=\sqrt{f_0A(\!t_i\!)T}$.
This causes an error, $(\Delta t_0)_i$,
in the determination of $t_0$ from a single-bin measurement given by
\begin{equation}
(\Delta t_0)_i =
(\Delta N)_i \left[{f_0T {dA(\!t_i\!)\over dt}}\right]^{-1}
= \sqrt{A(\!t_i\!)\over f_0T}\,\left({dA(\!t_i\!)\over dt}\right)^{-1}.
\end{equation}
Since the photon statistics errors for the different bins
are uncorrelated, the total error resulting from making
measurements of this type
for each bin on the entire light curve is given by
\begin{equation}
(\Delta t_0)^2=\left[{\sum_i\left({1\over\Delta t_i}\right)^2}\right]^{-1}
=\left[{f_0\int_{-\infty}^{\infty}
{{1\over A}\left({dA\over dt}\right)^2\,dt}}\right]^{-1},
\label{D1}
\end{equation}
where in the final expression we have taken the limit $T\rightarrow 0$.
By similar reasoning,
the answer to the second question is that
the error due to photon statistics in measuring the maximum amplification,
for given $\omega$ and $t_0$,
is
\begin{equation}
(\Delta A_m)^2=
\left[{f_0\int_{-\infty}^{\infty}
{\left({A^2-1\over{A_m}\!^2-1}\right)^{3}
{1\over A}\,dt}}\right]^{-1}.
\label{D2}
\end{equation}
The integrands of eqs.~\ref{D1} and~\ref{D2} are plotted in Fig.~\ref{fig1},
for the light curve corresponding to the $A=7$ event observed by
the MACHO group.~\cite{MACHO}

Equations~\ref{D1} and~\ref{D2} simplify considerably for events where
$A_m$ is large -- as will be the only case of interest here -- since then
eq.~\ref{Acurve} is well approximated by $A(u) = 1/u$ over the
relevant portion of the light curve. Substituting
$A(u) = 1/u$
into eqs.~\ref{D1} and~\ref{D2} and performing the integrals, we obtain,
for large amplification events,
\begin{equation}
(\Delta t_0)^2 = {3 \over 2f_0 \omega {A_m}\!^2}
\label{D1'}
\end{equation}
and
\begin{equation}
(\Delta A_m)^2= {3 \omega {A_m}\!^2 \over 4f_0}.
\label{D2'}
\end{equation}
In the large amplification approximation we find that the
the integrand of eq.~\ref{D1} has a double peak at times $t_P$ given by
\begin{equation}
t_P = t_0 \pm  {\sqrt{2/3} \over \omega A_m},
\end{equation}
where the width of each peak is of order ${1/ \omega A_m}$.
The integrand of eq.~\ref{D2} has a single peak at
$t = t_0$, with a width also of order ${1/ \omega A_m}$.

Equations~\ref{D1} and~\ref{D2}, and their approximations~\ref{D1'}
and~\ref{D2'}, provide lower bounds on the
errors arising in measurements of $t_0$ and $A_m$ by a single telescope.
Clearly, useful information about the MACHO can be extracted from
parallax  observations by two telescopes
only if the errors in measuring $t_0$ and/or $A_m$
by the individual telescopes are small compared with the differences,
$\delta t_0$ and $\delta A_m$, occurring in
the values of these parameters between
the two telescopes. If we have two (inertial) telescopes separated
by a transverse
distance $D$, then the shift in the central time between the telescopes
is given by~(see footnote 1)
\begin{equation}
\delta t_0 = {D\cos\theta\over\omega\tilde r_e},
\label{dt}
\end{equation}
where $\theta$ is the magnitude of the angle between $\tilde{\bf v}$
and the
line joining the telescopes.
Similarly, the shift in $\beta$ is given
by
\begin{equation}
\delta\beta={D\sin\theta\over\tilde r_e}.
\end{equation}
In the large amplification approximation (where $A_m = 1/ \beta$) this
yields
\begin{equation}
\delta A_m = -{D\sin\theta {A_m}\!^2 \over\tilde r_e}.
\label{dAm}
\end{equation}
Thus, the appropriate ``figures of merit" for parallax measurements
-- assumed to be limited only by photon statistics --
for large amplification events are
\begin{equation}
\left({\Delta t_0 \over \delta t_0}\right)^2 =
{3 \tilde{v}^2 \over 2f_0 \omega {A_m}\!^2 D^2 \cos^2\theta}
\label{F1}
\end{equation}
and
\begin{equation}
\left({\Delta A_m \over \delta A_m}\right)^2 =
{3 \tilde{v}^2 \over 4f_0 \omega {A_m}\!^2 D^2 \sin^2\theta}.
\label{F2}
\end{equation}
Remarkably, apart from the replacement of $\cos^2\theta$ by $\sin^2\theta$,
the right sides of eqs.~\ref{F1} and~\ref{F2} differ only by a
factor of $2$. This means that -- apart from different
angular sensitivities -- the time delay
and amplification differences are essentially equally useful
for measuring $\tilde{\bf v}$.

Solving eqs.~\ref{dt} and~\ref{dAm} for $\tilde{v} = \omega \tilde r_e$,
we obtain
\begin{equation}
\tilde{v}^2 = D^2 \omega^2 {A_m}\!^4
\left[{(\delta A_m)^2 + \omega^2 {A_m}\!^4 (\delta t_0)^2}\right]^{-1}.
\label{vtilde}
\end{equation}
For the purpose of making an order of magnitude estimate, we shall take
the errors caused by photon statistics in the determination of $t_0$
and $A_m$ as being independent, and
given (for each telescope) by eqs.~\ref{D1'} and~\ref{D2'} respectively. The
resulting error in the determination of the magnitude
of $\tilde{\bf v}$ from a parallax measurement by two inertial
telescopes is given by
\begin{equation}
{\Delta \tilde v \over \tilde v}= {\tilde v \over D}
\left[{3 \over 2f_0 \omega {A_m}\!^2}\right]^{1/2} \sqrt{1 + \cos^2 \theta}.
\label{Dv}
\end{equation}

Quite valuable information would be obtained from a parallax
measurement if it were able to determine $\tilde{v}$ accurately
enough to distinguish between a disk and halo population. As a
rough criterion for this, we require that one be able to make
a $\sim 50\%$ measurement of $\tilde{v}$
when $\tilde{v} \sim 80\,\rm km/sec$.
Let us estimate the conditions which must be satisfied by a lensing event
in order for this to occur. We consider a typical lensed star with
an unamplified magnitude of V=19. For a $2.5\,\rm m$ telescope and an
observing bandwidth of $1000$~\AA, this
corresponds to $f_0=10^3\,\rm photons/sec$.
For the purpose of making an estimate, we assume that two
$2.5\,\rm m$ telescopes are dedicated ``full time"  to the task of observing
the MACHO microlensing event
for, say, $3$ or $4$ nights near maximum amplification.
The effective photon flux is reduced
in the above formulas by a factor of $3$
to account for an $8$ hour per night observing period, and by another factor
of about $2$ to account for our
observing only near maximum amplification. (Since the
width of the peaks of the integrands in eqs.~\ref{D1} and~\ref{D2} vary
as ${1/ \omega A_m}$, a somewhat larger reduction would have to be
made for relatively low amplification or long timescale events.)
Choosing $D$ to be an Earth radius ($6,000\,\rm km$) and
taking $\cos^2 \theta = 1/2$, we find that
to distinguish between disk and halo populations in our two-telescope
idealized parallax experiment,
we need \footnote{
Note that the lower limit for $\omega {A_m}\!^2$ in eq.~\ref{estimate}
scales with $D$ as $D^{-2}$.
Hence, if the second telescope were placed in solar
orbit at a distance of $1\,\rm Au$ from Earth, one would
decrease this limit by
a factor of $5\times10^8$
(although it should be
remembered that eq.~\ref{estimate}
applies only in the high amplification
approximation). Thus, even taking into account the much smaller area
of an orbiting telescope, we see that photon counting statistics would
not impose any significant limitations on parallax measurements using
a telescope in solar orbit.}
\begin{equation}
\omega {A_m}\!^2 \gta 1\,\rm day^{-1}.
\label{estimate}
\end{equation}
Thus, for a typical value of $\omega = .05\,\rm day^{-1}$, we see that
the errors resulting from photon statistics alone will limit useful
measurements to the cases where $A_m \gta 5$. For Earth based
measurements, this limit could be
improved only by use of more (or larger) telescopes.
Note that since, for any given MACHO mass and velocity,
the probability of an event having impact parameter
less than $b$ is proportional to $b$,~\cite{Griest}
and since for large $A_m$ we have
$A_m \simeq 1/\beta$, it follows that about $20\%$
of all events will have $A_m \geq 5$ (where an ``event" is defined
by the criterion that $\beta \leq 1$). Most importantly, the
imposition of a selection criterion on $A_m$ alone randomly samples
the MACHO events, i.e., it does not bias the class of all
MACHO events with respect to any MACHO properties.
However, a restriction to events within
a range of timescales can introduce biases.

Our analysis above has ignored the rotation of the Earth. In fact, the
Earth's rotation is helpful in that it enables one, in principle,
to do parallax
measurements with a single telescope (since it provides an effective
baseline of order an earth radius) -- although, undoubtedly, one
would want to do the measurements with at least two telescopes
to improve statistics and reduce systematic errors.\footnote{
Note that with two telescopes separated by latitude on
a rotating Earth, one has effective baselines in both latitude and
longitude. Consequently one could, in principle, measure both components
of $\tilde{\bf v}$ by making use only of the information
contained in $\delta A_m$. Since, when suitably normalized,
the peak of the $(\Delta A_m)^2$ integrand tends to be higher and sharper
than the double peak of the $(\Delta t_0)^2$ integrand (see Fig.~1),
this suggests that the best
strategy for parallax observations might be to have many large telescopes
(separated in latitude) observe the event for one or two nights near
maximum amplification, rather than have one or two telescopes observe the
event for many nights.}
Nevertheless,
one might worry that the estimates used in deriving eq.~\ref{Dv} could be
altered by effects of the Earth's rotation. In addition,
the estimates of $\Delta t_0$ and $\Delta A_m$ in
eqs.~\ref{D1} and~\ref{D2} may have been
optimistic in that they were derived
under the assumption that all the parameters except the one in question
were known exactly. To check the validity of eq.~\ref{Dv}, we have written
a Monte Carlo code in which one or more telescopes were placed at
various geographic locations on a rotating Earth, each observing
for $8$ hours per night for several nights. The theoretical
light curve for
each telescope was then binned into 0.2 hour ``observations", and each
observation was then randomly altered by the appropriate
Gaussian $\sqrt N$ photon statistics errors. We then fit the resulting
observations with the six free parameters
$\mbox{\boldmath $\beta$}$, $\mbox{\boldmath $\omega$}$,
$t_0$, $\tilde r_e$.
Repeating this for a large number of trials, we arrived
at numerical estimates for $\Delta t_0$, $\Delta A_m$, and
$\Delta\tilde{\bf v}$. These numerical estimates were compared
with the corresponding analytic estimates, obtained by utilizing appropriate
``effective baselines" and modifying the range of the integrals in
eqs.~\ref{D1} and~\ref{D2} to cover only the period
of time during which the telescopes actually observed. The agreement between
the numerical and analytic estimates was very good (within a factor of $2$
in essentially all cases tried). Thus, we are confident that the formulas
given above provide reliable order of magnitude estimates of the
limitations imposed by photon statistics errors on parallax
measurements of MACHOs.

The key issue regarding the feasibility of Earth-based parallax
measurements of
high amplification events ($A_m\gta 5$)
is whether all other possible sources of error
can be reduced to below that of the photon statistics error computed
above.  The challenges to doing this are quite
formidable. However, it should be noted that some of the conditions
present in a microlensing event
are extremely favorable for doing accurate photometry. In
particular, the brightness change of the lensed star is completely
achromatic, and the required brightness measurements are differential
both with respect to time and with respect to the background stars in
the field. Since the lensing events are observed in crowded star fields,
there should be no difficulty finding an appropriate ``mix" of nearby
stars to match the color of the lensed star. Thus, it should be possible
to eliminate the effects of time dependent and color dependent
changes in the absorption by the atmosphere. Unfortunately,
crowded fields also exacerbate what is probably the most serious source of
error: the time dependence of seeing conditions, which causes
a spread in the
stellar images. With the high magnifications attainable by large
telescopes (which would have to be used for the parallax measurements
in any case), it
should be possible to accurately measure, and correct for,
the effects of seeing conditions on stars in a field.
However, we must leave a detailed analysis of sources of error
and other practical limitations to those more experienced than we
in such matters.

We wish to thank Don Lamb for pointing out that an argument given
to him by one of us to argue against the possibility of gamma ray bursts being
lensing events\footnote{
The argument is simply that if gamma-ray bursts
corresponded to our passage through caustics resulting from the gravitational
lensing of a distant, continuous source,
then the time delay between receipt of a burst by
two detectors would be given by a special relativistic version of
eq.~\ref{dt}, corresponding to an effective velocity smaller than $c$. In
addition, the direction of the burst as inferred from timing measurements
would not coincide with the direction of propagation of the gamma rays.}
could potentially be applied to the study of
MACHOs. We wish to thank Rich Kron for reading the manuscript and for
several very useful discussions concerning observational issues.
We thank Andrew Gould for pointing out an error in the original version
of this manuscript.
This research was supported in
part by National Science Foundation
grant PHY-9220644 to the University of Chicago.
D.H. acknowledges support from an Air Force
Laboratory Graduate Fellowship.

\begin{figure}
\caption[figure1]{
The integrands of eqs.~\ref{D1} and~\ref{D2},
plotted for the first event reported by the MACHO group
($A_m=6.86$, $1/\omega=16.95$).~\cite{MACHO}
The $(\Delta t_0)^2$ integrand has been multiplied by $\omega^{-2}
 {A_m}\!^{-4}$
in order to make it dimensionless; this also
weights it appropriately
relative to the $(\Delta A_m)^2$ integrand with respect to utilizing $t_0$
and $A_m$ to obtain information about the velocity of the MACHO
(see eq.~\ref{vtilde}
below). From the figure, we see that essentially all of the
information concerning $A_m$ can be extracted from observations
made during a period of a few days
near maximum amplification, whereas the information concerning $t_0$
is peaked (also with a spread of a few days)
two days before and after maximum amplification.}
\label{fig1}
\end{figure}

\end{document}